\documentstyle[prl,aps,twocolumn,epsfig]{revtex}

\begin{document}
\draft
\title
{Avoided Band Crossings:\\ Tuning Metal-Insulator Transitions in
Chaotic Systems} 
\author{R. Ketzmerick$^{1,2}$,
K. Kruse$^{2}$, and
T. Geisel$^{1,2}$}

\address{
$^1$Institute for Theoretical Physics, 
University of California Santa Barbara, CA 93106, USA \\
$^2$Max-Planck-Institut f\"ur Str\"omungsforschung und Institut f\"ur
Nichtlineare Dynamik der \\ Universit\"at G\"ottingen, 
Bunsenstra{\ss}e 10, D-37073 G\"ottingen, Germany$^*$\medskip\\
\parbox{14cm}{\rm 
We show that avoided crossings of energy bands may give rise to a variety of
phenomena such as transitions from metal to insulator and vice versa,
changes in localization lengths, and changes in the fractal 
dimension of energy spectra. We explain the occurrence of these
phenomena in the kicked Harper 
model under classically chaotic conditions and predict them to 
occur in other systems.
}}

\maketitle
\narrowtext

Spatially periodic quantum systems give rise to energy {\em bands} and
extended eigenfunctions, according to Bloch's theorem. If this
periodicity is broken by a weak perturbation, e.g., by disorder or a
(magnetic) superlattice, the spectrum remains band-like. Typically, one
neglects the coupling of these "bands" and studies them in a one-band
approximation. Prominent examples from solid state physics for such one-band
models are the Anderson model for disordered 
systems~\cite{anderson} and the Harper model for Bloch electrons in a
magnetic field~\cite{peierls,harper},
which display such interesting properties as localized eigenfunctions
or metal-insulator transitions. As a function of an external parameter
some of these "bands" will move up in energy whereas others move down,
thus giving rise to band crossings. At these crossings the
coupling of bands can no longer be neglected and will lead to {\em
avoided} band crossings (ABCs) in analogy to the well-studied avoided
level crossings. The question then arises: Will these couplings of
bands and in
particular the ABCs change the properties derived in the simplifying
one-band models? 

A seemingly unrelated problem has occurred in the quantum chaos
literature, while studying the influence of classical chaos on
quasiperiodic quantum systems. Specifically, in the kicked Harper
model (KHM)~\cite{leboeuf1,lima,geisel,ketzmerick,artuso1,artuso2,guarneri,artuso3,roncaglia,leboeuf2,dana,borgonovi},
it was found numerically that in the regime, 
where the classical limit is chaotic, extended eigenfunctions show up
for parameters where exclusively localized ones were expected and vice
versa~\cite{lima,artuso2}. In addition, chaos was found to
change the fractal dimension of the spectrum~\cite{artuso1}. In the
absence of an explanation these puzzling phenomena have remained
mysterious, and one may wonder: How can classical chaos induce such
drastic changes in these quantum properties?

In fact, it will turn out here that the main impact of classical chaos
is to generate ABCs and thus this question for the quantum chaos
problem will be reduced to the previous question posed for the solid
state problems. In the main part of this paper we therefore analyze
avoided crossings of {\em one-band models}. Although simple
generalizations of avoided level crossings, these
ABCs unexpectedly generate transitions from localized to extended
eigenfunctions and vice versa and changes in the localization lengths
of eigenfunctions as well as in the fractal dimensions of spectra. We show
that ABCs cause these phenomena by effectively changing the parameters
of the one-band models involved. We will demonstrate that these
general findings can be applied to the KHM and thereby give the
first explanation for the phenomena that were observed numerically in
this model. Application to the old problem of Bloch electrons in a
magnetic field~\cite{peierls} leads to the prediction of dramatic
effects, e.g.,\ a wave packet spreading in one direction will, due to
ABCs, spread in the perpendicular direction. These effects give rise
to experimental consequences for transport measurements in
semiconductor nanostructures and will be described in detail
elsewhere~\cite{ketzmerick2}. In the future we expect applications to
coupled one-dimensional chains and many other systems. 

Let us start by studying an ABC when there is no perturbation and the
system is still periodic. Then each 
\begin{figure}[h]
\hspace*{0.5cm}
\epsfig{figure=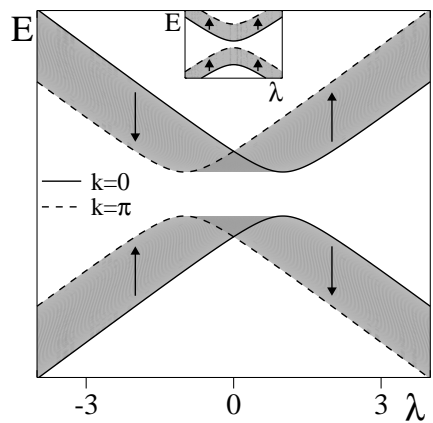,width=8.2cm}
\vspace*{-0.2cm}
\caption{An ABC with twist modeled by the Hamiltonian~(\ref{twist}),
with $E^\prime=-\cos k$, $E^{\prime\prime}=\cos k$, and $\epsilon = 0.6$.
The eigenvalues form two energy {\em bands} for the set of all Bloch
phases $k$. Arrows indicate in which direction the energies change as $k$ is
varied from $0$ to $\pi$. Solid and dashed lines represent the
eigenvalues for $k=0$ and $k=\pi$, resp. The inset shows an ABC without twist
for the case $E^\prime=E^{\prime\prime}=-\cos k$ with $\epsilon=1.4$.}
\end{figure}
\noindent band is described by a dispersion
$E(k)$ depending on the Bloch phase $k\in\left[-\pi,\pi\right]$. Now, 
an isolated avoided crossing of two bands can be modeled by the
Hamiltonian 

\begin{equation}\label{twist}
H = \left(\begin{array}{cc}
       E^\prime(k) + \lambda & \epsilon \\
       \epsilon & E^{\prime\prime}(k) - \lambda 
    \end{array}\right),
\end{equation}

\noindent where $\epsilon$ is the coupling parameter. For any fixed
Bloch phase $k$, $H$ simply models an avoided level crossing. When looking at 
the entire band there are two cases of interest: If the
dispersion relations of the two bands have the same general dependence
on $k$, namely $\frac{dE}{dk}>0$ for $k\in\left[ 0,\pi\right]$, the resulting
spectrum resembles that of a broadened avoided level crossing (see
inset of Fig.~1). In 
the alternative case of opposite general dependencies on $k$, namely 
$\frac{dE^\prime}{dk}>0$ and $\frac{dE^{\prime\prime}}{dk}<0$ for
$k\in\left[ 0,\pi\right]$, each band is twisted (Fig.~1). 

Two subsequent ABCs with twist (see Fig.~2a) can be modeled by the
Hamiltonian 

\begin{equation}\label{doppeltwist}
H = \left(\begin{array}{ccc}
       E^+(k) +A-\lambda & \epsilon & 0 \\
       \epsilon & E^0(k)+\lambda & \epsilon \\
       0 & \epsilon & E^-(k)-A-\lambda
    \end{array}\right),
\end{equation}

\noindent where a rising band $E^0$ subsequently crosses two decreasing bands
$E^-$ and $E^+$ with offsets $\pm A$. For small coupling
$\epsilon$ the two twists of the intermediate band are well separated,
whereas for increasing 
$\epsilon$ they first come closer whereby the band flattens, which
will be important later on, and finally the twists
annihilate (Fig.~2). 

We now show the dramatic effect of ABCs in the presence of a weak
perturbation. This we model by the Hamiltonian of
Eq.~(\ref{doppeltwist}) with the bands
$E^{\pm,0}(k)$ replaced by Hamiltonians $H^{\pm,0}$, leading to the Hamiltonian

\begin{equation}\label{dreiketten}
H = \left(\begin{array}{ccc}
       H^+ +A-\lambda & \epsilon & 0 \\
       \epsilon & H^0 +\lambda & \epsilon \\
       0 & \epsilon & H^- -A-\lambda
    \end{array}\right).
\end{equation}

\noindent For simplicity, we will focus on tight-binding Hamiltonians
$H^{\pm,0} = \sum_nV^{\pm,0}_n\:a_n^\dagger a_n +
\sum_nt^{\pm,0}\,\left(a_{n+1}^\dagger a_n + a_{n-1}^\dagger a_n\right)$ 

\begin{figure}[h]
\hspace*{-1.1cm}
\epsfig{figure=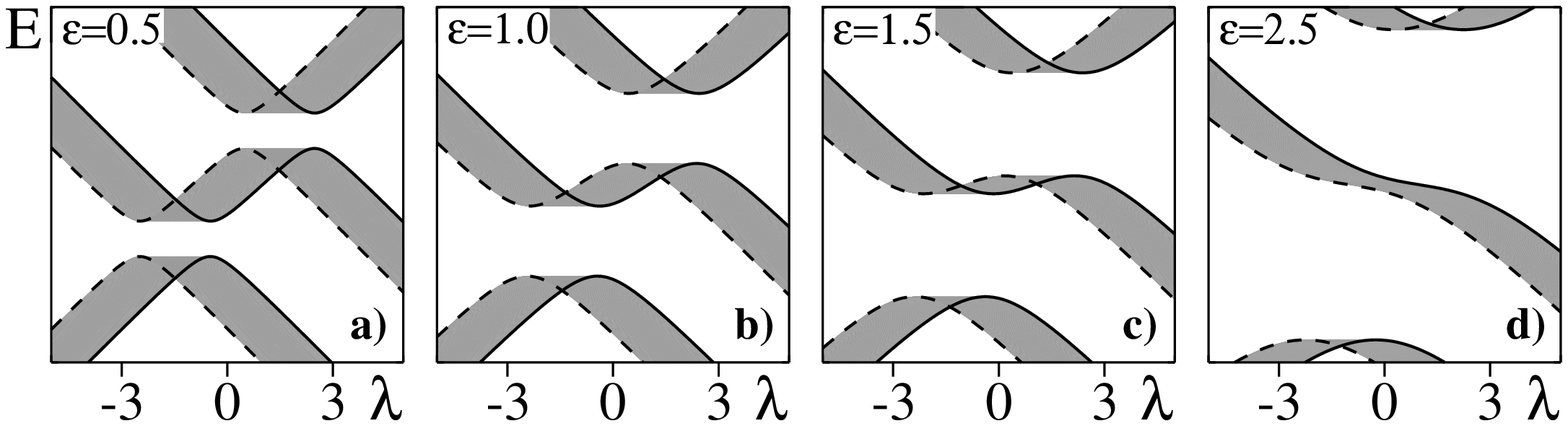,width=9.5cm}
\vspace*{-4.6cm}
\caption{Two subsequent ABCs with twist generated by the
Hamiltonian~(\ref{doppeltwist}) with $E^0(k)=-\cos k$, $E^\pm (k)=\cos
k$, and offset $A=3$ for increasing coupling $\epsilon$. For small
$\epsilon$ the twists are well 
separated in $\lambda$ (a). Upon increasing $\epsilon$ the two twists of the
intermediate band approach each other and
reduce its bandwidth (b,c). If $\epsilon$ is increased
further, the twists annihilate (d).}
\end{figure}
\begin{figure}
\hspace*{-1.2cm}
\epsfig{figure=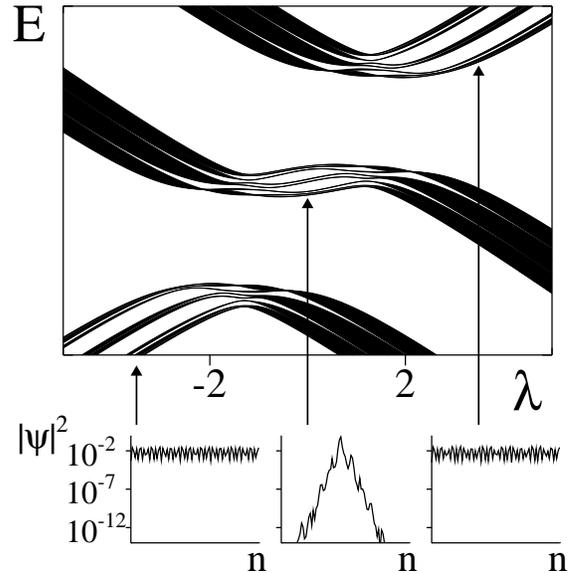,width=10cm}
\vspace*{0.3cm}
\caption{The spectrum of Eq.~(\ref{dreiketten}) where $H^0$ represents
the Harper 
model in the metallic regime ($V^0/t^0=1.2$, $\sigma=5/8$) and $H^\pm$ has
zero on-site 
potential and hopping terms $t^\pm=-t^0$ leading to twists. The
coupling parameter $\epsilon=1.5$ corresponds to Fig.~2c. At the
bottom typical
eigenfunctions are shown where one clearly sees a transition from extended
eigenfunctions in the uncoupled regime ($|\lambda |\gg1$) to localized
eigenfunctions in the regime between the twists ($\lambda=0$).}
\end{figure}
\noindent on one-dimensional lattices with {\em spatially varying} on-site
potential $V_n^{\pm,0}$ and with
hopping terms $t^{\pm,0}$. For zero on-site potentials ($V_n^{\pm,0}\equiv 0$)
Eq.~(\ref{dreiketten}) reduces to Eq.~(\ref{doppeltwist}) studied in Fig.~2.

As a first numerical example we assume $H^0$ according to Harper's
model for Bloch electrons in a magnetic field~\cite{harper} in the 
metallic regime, i.e.,\ we take the on-site potential $V^0_n=V^0\cos
(2\pi \sigma n)$ and $V^0/t^0<2$, where $\sigma$ is the magnetic flux
per unit cell. For $H^\pm$ we assume vanishing on-site potential
$V^\pm_n\equiv 0$
and, in order to generate twists, hopping terms $t^\pm =-t^0$. Fig.~3
corresponds to Fig.~2c, where the three tight-binding bands now show fine 
structure, while on a coarse scale they display the same twists and
flattening as in Fig.~2c. For large $\left|\lambda\right|$ the
individual spectra of 
$H^\pm$ and $H^0$ are uncoupled and all eigenfunctions are {\em
extended}. Surprisingly, in the $\lambda$ region between the twists
the intermediate band has {\em localized} eigenfunctions only. 
If for $H^0$ we assume the Anderson model for disordered
systems~\cite{anderson}, where 
$V_n^0$ is a uniform random variable in the interval
$\left[-V^0/2,V^0/2\right]$, for $\lambda=0$ we find a decreasing
localization length of eigenfunctions as the coupling parameter
$\epsilon$ is increased (Fig.~4a, squares).
As a third example for $H^0$ we assume the Fibonacci chain model
\cite{fibonacci}, a 
one-dimensional model for quasicrystals with potential terms taking
on the values $V^0$ and $-V^0$ according to a Fibonacci
sequence. Remarkably, at $\lambda=0$ increasing
$\epsilon$ lowers the fractal dimension of the energy spectrum
(Fig.~4b, squares).
These three systems show 
\begin{figure}
\vspace*{-0.2cm}
\hspace{-.8cm}
\epsfig{figure=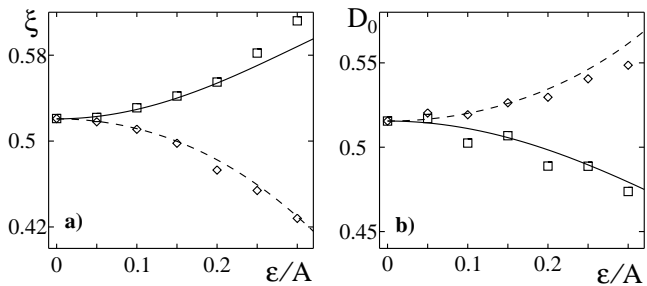,width=8.8cm}
\vspace*{-2.0cm}
\caption{\noindent a) Average inverse participation ratio
$\xi=\sum_n \left|\psi_n\right|^4$ of the eigenfunctions $\psi$ of the
intermediate band of Eq.~(\ref{dreiketten}) for $\lambda=0$, $A=10$, and
different values of $\epsilon$. Here $H^0$ represents the Anderson model
($V^0= 5$, $t^0=1$) and $H^\pm$ has zero on-site potential and
hopping terms $t^\pm=-1$ (squares) and $t^\pm=1$ (diamonds), resp.
Solid ($t^\pm=-1$) and dashed ($t^\pm=1$) lines are the results for an
isolated Anderson model with the effective terms $\bar{V}_n^0$ and
$\bar{t}^{\,0}$ of perturbation theory
(Eqs.~(\ref{Vt_eff_mit_twist}) and~(\ref{Vt_eff_ohne_twist})). An
increasing (decreasing) inverse participation ratio corresponds to a
decreasing (increasing) localization length. \\ 
\noindent b) Same as (a) showing the fractal (box-counting) dimension
for $H^0$ representing the Fibonacci model ($V^0= 1.5$, $t^0=1$, $A= 100$).}
\end{figure}
\noindent the opposite of what one would
expect naively: The 
coupling of a given tight-binding model ($H^0$) to models
having {\em extended} eigenfunctions ($H^\pm$) in these three
cases leads to {\em localization}, {\em reduced} localization lengths, and
{\em reduced} fractal dimensions, respectively. How is that possible?

One can try to understand these effects of ABCs
intuitively: The reduced width of the intermediate band in Fig.~2c
resulting from
subsequent ABCs with twists corresponds to reduced hopping terms in a
tight-binding Hamiltonian~\cite{holthaus}. More generally, one may 
expect that the 
ratio of potential terms to hopping terms is increased by these ABCs
with twist. This would naturally explain the observed results, as for
increasing $V^0/t^0$ the Harper model makes a transition from metal to
insulator~\cite{harper}, the Anderson model has a decreasing
localization length~\cite{anderson},
and the energy spectrum of the Fibonacci model has a decreasing
fractal dimension~\cite{hiramoto}.

Further insight can be obtained from a perturbative treatment for
$\lambda=0$. By a similarity transformation we obtain
$\bar{H}=S^{-1}HS$, where

\begin{equation}\label{S}
S = \left(\begin{array}{ccc}
       \left( 1-\epsilon^2/2A^2\right) & -\epsilon/A & \epsilon^2/2A^2 \\
       \epsilon/A & \left( 1-\epsilon^2/A^2\right) & -\epsilon/A \\
       \epsilon^2/2A^2 & \epsilon/A & \left( 1-\epsilon^2/2A^2\right)
    \end{array}\right).
\end{equation}

\noindent This choice of $S$ leads to renormalized Hamiltonians
$\bar{H}^{\pm,0}$ analogous to Eq.~(\ref{dreiketten}) that
are decoupled up to order $\epsilon^2/A^2$ for
$|V_n^{\pm,0}|,|t^{\pm,0}|\ll A$. The Hamiltonian $\bar{H}^0$
describing the intermediate band is still tridiagonal with on-site
potential $\bar{V}_n^0$ and hopping terms $\bar{t}^{\,0}$ up to order
$\epsilon^2/A^2$ given by

\begin{equation}\label{Vt_eff_mit_twist}
  \begin{array}{lcl}
    \bar{V}_n^0 & = & V^0_n\left(1-2\epsilon^2/A^2\right) + 
                      \left( V_n^+ +V_n^- \right)\epsilon^2/A^2 \\
    \bar{t}^{\,0}   & = & t^0\left(1-4\epsilon^2/A^2\right).
  \end{array}
\end{equation}

\noindent For the systems studied above (where $V_n^+\equiv
V_n^-\equiv 0$) we find that the ratio $V_n^0/t^0$ is effectively {\em
increased} by the factor $\left(1+2\epsilon^2/A^2\right)$. This
substantiates the intuitive explanation of the preceding paragraph.
E.g.,\ in the Harper model ABCs with twist may induce a transition from
extended states ($V^0/t^0<2$) to localized
states ($\bar{V}^0/\bar{t}^0>2$).
For two subsequent ABCs without twist, i.e.\ when the hopping terms $t^\pm$ and
$t^0$ have the same sign, the above perturbative treatment yields

\begin{equation}\label{Vt_eff_ohne_twist}
  \begin{array}{lcl}
    \bar{V}_n^0 & = & V^0_n\left(1-2\epsilon^2/A^2\right) + 
                      \left( V_n^+ +V_n^- \right)\epsilon^2/A^2 \\
    \bar{t}^{\,0}   & = & t^0
  \end{array}
\end{equation}

\noindent instead. Thus, without twists the ratio $V_n^0/t^0$ effectively {\em
decreases} 
by the factor $(1-2\epsilon^2/A^2)$. In this case we therefore expect
the opposite 
effects on eigenfunctions and spectrum. Using the renormalized terms
of Eqs.~(\ref{Vt_eff_mit_twist}) and (\ref{Vt_eff_ohne_twist}) we can
compute the localization length for an isolated Anderson model
and the fractal dimension $D_0$ for an isolated Fibonacci
band. Comparison with the numerical results in Fig. 4a) and b) gives a
good agreement and thus confirms our explanation in terms of
renormalized tight-binding parameters. In contrast to the case of ABCs
with twist, here an intuitive understanding is missing, and one has to
rely on the above perturbative treatment.
The phenomena associated with ABCs analyzed above
also do occur when there is just a
single ABC, although in a more complex way. Thus
they should be generic for all models with ABCs.

Now we will apply these general results to the well-studied kicked
Harper model (KHM) 
\cite{leboeuf1,lima,geisel,ketzmerick,artuso1,artuso2,guarneri,artuso3,roncaglia,leboeuf2,dana,borgonovi}.
This model combines the quantum mechanical properties of Harper's 
equation with an onset of chaos in classical phase space. It
shows similar phenomena~\cite{lima,artuso1,artuso2}, as those
discussed for ABCs, but they have
remained unexplained so far.
The KHM is given by the Hamiltonian

\begin{equation}\label{khm}
  H = L\cos p + K\cos x \sum_n\delta\left( t-n\right)
\end{equation}

\noindent with $p=-i\hbar\partial/\partial x$ and an effective
$\hbar$. For small $K$ and $L$ its spectrum is close to the Harper
spectrum for $\sigma=\hbar /2\pi$ and $V^0/t^0=2K/L$ up to a scaling by a
factor proportional to 
$K$~\cite{geisel} (Fig.~5). In particular, the spectrum clusters into subbands
each of which may be described by a Harper-like
model~\cite{wilkinson}. Upon increasing the parameters $K$ and $L$
these subbands make band crossings. As soon as the classical limit is
non-integrable (i.e., mixed or fully chaotic) such crossings are
avoided. Therefore our general findings on ABCs apply and one may
expect all the phenomena discussed above to occur. Indeed, this is the
case. For example, for the parameters of Fig.~5 transitions from
extended to localized eigenfunctions can
\begin{figure}
\vspace*{-0.1cm}
\hspace*{-2.2cm}
\epsfig{figure=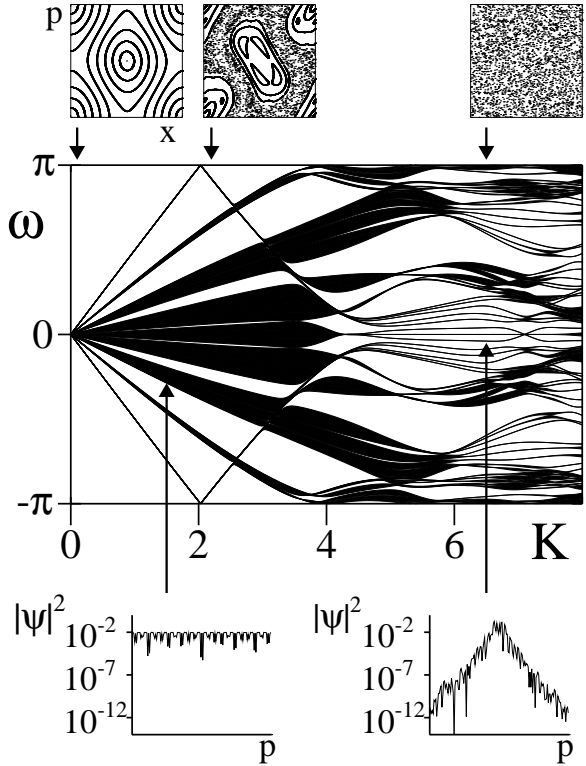,width=13.6cm}
\vspace*{0.3cm}
\caption{Transitions of the quasienergy spectrum of the kicked Harper
model for increasing kicking strength $K$ and $L/K=4/7$ as the
classical phase space becomes increasingly chaotic (insets on
top). Here $\hbar/2\pi=8/61$ is a rational approximant of $1/(6+\sigma_G)$
where $\sigma_G=1.618...$ is the golden mean and the Bloch phase in 
$p$-direction is varied. While for small $K$ and
$L$ all eigenfunctions are extended as expected (a typical one is
shown at the bottom), for a chaotic phase space and stronger kicking
$K$ one simultaneously finds localized and extended states (bottom)
depending on whether or not ABCs and twists have led to flattened bands.}
\end{figure}
\noindent be observed as were
numerically found in Refs.~\cite{lima,artuso2}. They occur along with
ABCs, however, in a more complex way than in the simple model system
of Eq.~(\ref{dreiketten}). Tuning these metal-insulator
transitions necessarily leads to states right at the transition point,
which, e.g.,\ in the Harper model correspond to a singular continuous
spectrum. We therefore also understand the findings of Borgonovi and
Shepelyansky~\cite{borgonovi}, that for the dual parameters of Fig.~5 at 
$K=4$ and $L=7$
part of the spectrum seems to be singular continuous~\cite{remark}. 
Similarly, the increase of the fractal dimension of the spectrum for
increasing $K=L$~\cite{artuso1} can be understood on the basis of ABCs
as analyzed above in the Fibonacci model (Fig.~4b).

This work was carried out in part during two of the authors' stay at
the Institute for Theoretical Physics, Santa Barbara. T.~G. and
R.~K. gratefully acknowledge the hospitality of the ITP and its members.
This work was supported by the NSF under Grant No. PHY94-07194 and in
part by the Deutsche Forschungsgemeinschaft.


\begin{references}
\vspace*{-1cm}
\bibitem[*]{address}
Present and permanent address.

\bibitem{anderson}
P.~W.~Anderson, Phys.\ Rev.\ {\bf 109}, 1492 (1958).

\bibitem{peierls}
R.~Peierls, Z.\ Phys.\ {\bf 80}, 763 (1933).

\bibitem{harper}
P.~G.~Harper, Proc.\ Phys.\ Soc.\ {\bf A68}, 874 (1955);
M. Ya. Azbel, Sov.\ Phys.\ JETP {\bf 19}, 634 (1964);
D.~R.~Hofstadter, Phys.\ Rev.\ B\ {\bf 14}, 2239 (1976);
S.~Aubry and G.~Andr\'e, Ann.\ Israel\ Phys.\ Soc.\ {\bf 3}, 133
(1980).

\bibitem{leboeuf1}
P.~Leb{\oe}uf, J.~Kurchan, M.~Feingold, and D.~P.~Arovas, Phys.\ Rev.\
Lett.\ {\bf 65}, 3076 (1990). 

\bibitem{lima}
R.~Lima and D.~Shepelyansky, Phys.\ Rev.\ Lett.\ {\bf 67}, 1377
(1991).

\bibitem{geisel}
T.~Geisel, R.~Ketzmerick, and G.~Petschel, Phys.\ Rev.\ Lett.\ {\bf
67}, 3635 (1991). 

\bibitem{ketzmerick}
R.~Ketzmerick, G.~Petschel, and T.~Geisel, Phys.\ Rev.\ Lett.\ {\bf
69}, 695 (1992). 

\bibitem{artuso1}
R.~Artuso, G.~Casati, and D.~Shepelyansky, Phys. Rev. Lett. {\bf 68},
3826 (1992). 

\bibitem{artuso2}
R.\ Artuso, F.~Borgonovi, I.~Guarneri, L.~Rebuzzini, and G.~Casati,
Phys.\ Rev.\ Lett.\ {\bf 69}, 3302 (1992). 

\bibitem{guarneri}
I.~Guarneri and F.~Borgonovi, J.\ Phys. {\bf A26}, 119 (1993).

\bibitem{artuso3}
R.~Artuso, G.~Casati, F.~Borgonovi, L.~Rebuzzini, and I.~Guarneri,
Int.\ J.\ Mod.\ Phys.\ B {\bf 8}, 207 (1994). 

\bibitem{roncaglia}
R.~Roncaglia, L.~Bonci, F.~M.~Izrailev, B.~J.~West, and P.~Grigolini,
Phys.\ Rev.\ Lett.\ {\bf 73}, 802 (1994).

\bibitem{leboeuf2} 
P.~Leb{\oe}uf and A.~Mouchet, Phys.\ Rev.\ Lett.\ {\bf 73}, 1360
(1994).

\bibitem{dana}
I.~Dana, Phys.\ Rev.\ Lett.\ {\bf 73}, 1609 (1994).

\bibitem{borgonovi}
F.~Borgonovi and D.~Shepelyansky, Europhys. Lett. {\bf 29}, 117 (1995).

\bibitem{ketzmerick2} R.~Ketzmerick, K.~Kruse, D.~Springsguth, and
T.~Geisel, to be published. 

\bibitem{fibonacci}
M.~Kohmoto, L.~P.~Kadanoff and C.~Tang, Phys.\ Rev.\ Lett.\ {\bf 50},
1870 (1983); S.~Ostlund, R.~Pandit, D.~Rand, H.~J.~Schellnhuber, and
E.~D.~Sigga, Phys.\ Rev.\ Lett.\ {\bf 50}, 1873 
(1983).

\bibitem{holthaus}
A reduction of hopping terms due to a band collapse in spatially
periodic systems subjected to a time periodic field was studied in
M.~Holthaus,
G.~Ristow, and D.~W.~Hone, Phys.\ Rev.\ Lett.\ {\bf 75}, 3914 (1995);
K.~Drese and M.~Holthaus, Phys.\ Rev.\ Lett.\ {\bf 78}, 2932 (1997).

\bibitem{hiramoto}
H.~Hiramoto and S.~Abe, J. Phys. Soc. Jpn. {\bf 57} (1), 230 (1988);
T.\ Geisel, R.\ Ketzmerick, and G.\ Petschel, in {\em Quantum Chaos -
Theory and Experiment}, edited by P.\ Cvitanovic, I.\ C.\ Percival,
and A.\ Wirzba (Kluwer, Dordrecht, 1992), p.43.

\bibitem{wilkinson}
M.~Wilkinson and R.~J.~Kay, Phys.\ Rev.\ Lett.\ {\bf 76}, 1896 (1996).

\bibitem{remark}
It actually turns out that every state passes the transition point for
a different value of $K$ such that for fixed $K$ the spectrum has
no singular continuous component. We were able to verify this using
a new efficient method
for diagonalizing kicked quantum systems. R.~Ketzmerick, K.~Kruse, and
T.~Geisel to be published.

\end{references}
\end{document}